\documentclass[pra,twocolumn,floatfix]{revtex4-1}

\usepackage[utf8]{inputenc}
\usepackage[backref=none]{hyperref}
\hypersetup{
    colorlinks=true,          
    linkcolor=blue,           
    citecolor=blue,           
    urlcolor=green,           
}

\usepackage{amsmath}
\usepackage{amssymb}
\usepackage{multirow}
\usepackage{graphicx}
\graphicspath{ {images/} }
\usepackage{array}
\usepackage{upgreek}

\makeatletter

\begin{document}

\title{Efimov energy level rebounding off the atom-dimer continuum}

\author{Yaakov Yudkin}
\author{Roy Elbaz}
\author{Lev Khaykovich}

\affiliation{Department of Physics, QUEST Center and Institute of Nanotechnology and Advanced Materials, Bar-Ilan University, Ramat-Gan 5290002, Israel}

\date{\today}

\begin{abstract}

The Efimov effect, with its ladder of weakly bound three-atomic molecules, poses intriguing questions in the theoretically controversial and experimentally demanding regime of merging of the first excited Efimov energy level with the atom-dimer continuum.
Using an original experimental technique, where a coherent superposition state of diatomic and triatomic molecules is utilized, we investigate this regime and reveal a striking behavior:
Instead of merging with the atom-dimer continuum the trimer energy level rebounds from it and becomes a deeper bound state again.
In addition, instead of a tangential approach between the two levels we observe a rather narrow resonance, providing a new challenge and guide for few-body theories to incorporate realistic interatomic potentials.

\end{abstract}

\maketitle

\section{Introduction}

The quantum few-body problem, in which a small number of particles is considered at length and energy scales for which quantum mechanics governs their behavior, is a cornerstone in the bottom-up approach to study the fundamental laws of physics~\cite{Kunitski15,Wang18,Liang18}.
Due to their superb versatility, ultracold atom systems, 
have proven to be one of the main experimental platforms for this research~\cite{Wenz13,Liu18,Guan19,Reynolds20}.
At the heart of the few-body problem lies the so-called Efimov effect which describes the formation of weakly bound triatomic molecules in the regime where the two-body interactions become extraordinarily large~\cite{Efimov70,Braaten&Hammer06,Greene17,Naidon17,D'Incao18}.
The growing evidence for the emergent three-body correlations in strongly interacting quantum gases strengthens the far-reaching role of the Efimov effect in bridging the few- and many-body physics~\cite{Levinsen15,Fletcher17,Klauss17,Sun17,Pierce19}.
Therefore, its thorough understanding over the entire parameter space is of paramount importance.

\begin{figure*}
\centering
\includegraphics[width=0.8\linewidth]{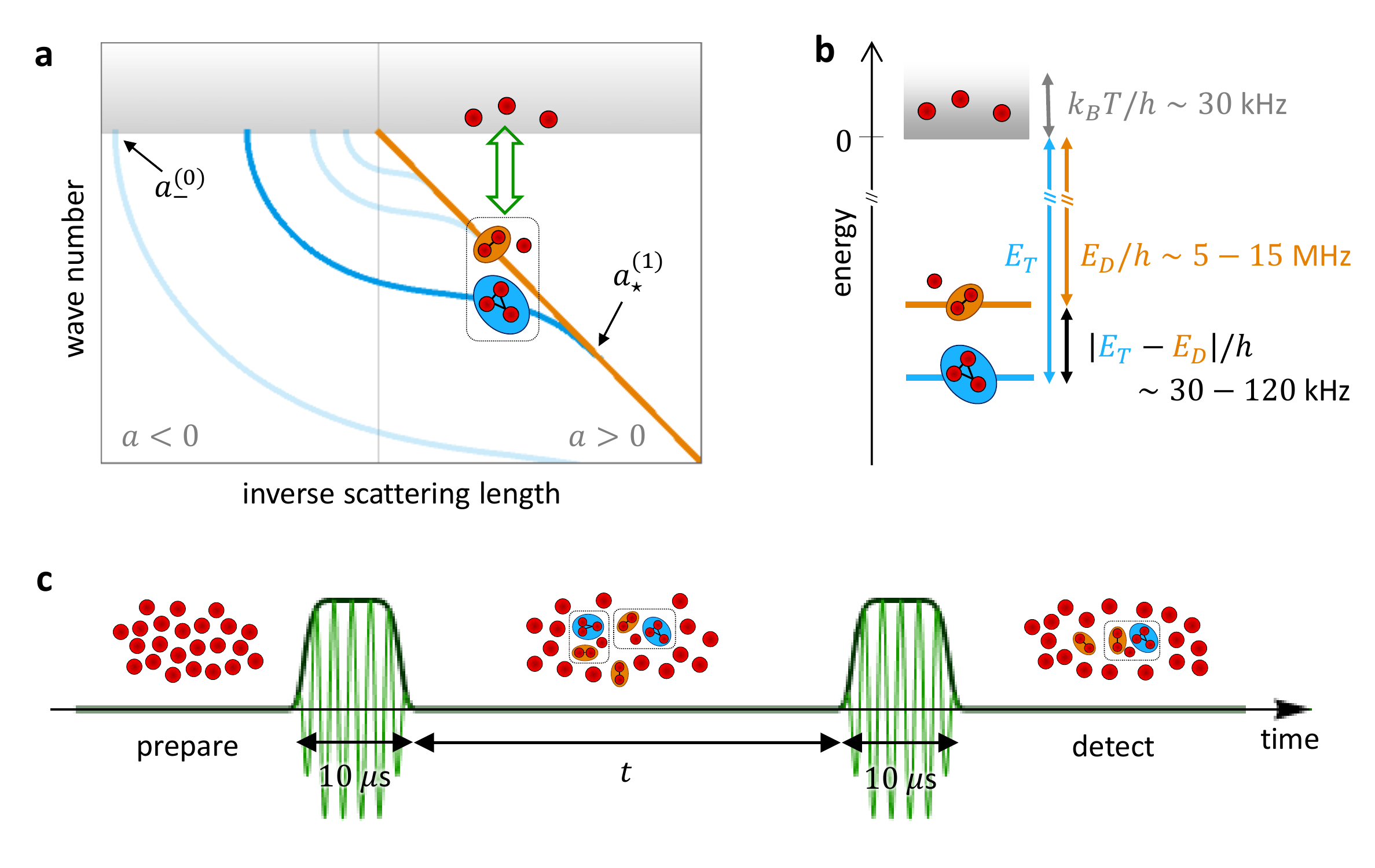}
\caption{\label{fig:illustration}
\textbf{Efimov scenario and experimental sequence.}
{\bf a.}
Universal theory for the two- (orange) and three-body (blue) bound state energies as a function of the inverse scattering length.
The green double arrow illustrates the coupling between the free atom continuum (shaded grey) and the DITRIS state generated by a short rf pulse.
{\bf b.}
Relevant energy scales.
{\bf c.}
Experimental Sequence: A fraction of the initially free atoms is converted to DITRIS states by the first pulse.
The second pulse, after a free evolution time $t$, attempts to cast them back onto the free atom continuum leading to interference of the number of atoms $N(t)$.}
\end{figure*}

The versatility of ultracold gases includes the tunability of the two-body interactions, characterized by the {\it s}-wave scattering length $a$, using Feshbach resonances~\cite{Chin10}.
Efimov physics is revealed when $|a|\gg r_{vdW}$ ($r_{vdW}$ is the length scale associated with the two-body van der Waals interaction potential) and among its peculiar features are a universal (i.e. independent of short-range details) discrete scaling of the binding energies $E_{T}^{(n)}$ ($n=0,1,2,\dots$ denotes the $n$-th trimer), namely $E_{T}^{(n)}/E_{T}^{(n+1)}\approx22.7^2$, and the fact that it remains bound for both positive and negative $a$, even though the conventional two-body bound state, the Feshbach dimer, only exists for $a>0$ (Fig.~\ref{fig:illustration}a).

When $|a|\gg r_{vdW}$, all three-body observables depend on two quantities: the scattering length $a$ and a three-body parameter which serves as a boundary condition to prevent the Thomas collapse~\cite{Thomas35}.
It is well established by now that the three-body parameter, represented by the scattering length $a=a_{-}^{(0)}$ at which the ground state Efimov trimer merges with the free atom continuum (for $a<0$, see Fig.~\ref{fig:illustration}a), is universally related to $r_{vdW}$ ($a_{-}^{(0)}=-9.7\,r_{vdW}$) for predominantly open-channel dominated ($s_{res}\gg1$) Feshbach resonances~\cite{Gross10,Berninger11,Wang12,Naidon14}.
This means that only the two-body quantities, $a$ and $r_{vdW}$, are necessary to calculate three-body observables and the notion of Efimov universality (sometimes called Efimov-van der Waals universality) reaches further than initially thought.
However, Efimov trimers in the vicinity of a closed-channel dominated Feshbach resonance ($s_{res}\ll 1$) do not follow this simple rule as was shown in recent precision measurements~\cite{Johansen17,Chapurin19}.
The common explanation is that they are more sensitive to finite-range corrections~\cite{Langmack18}.

On the positive scattering length side of the Efimov spectrum the available experimental data is sparse and inconclusive~\cite{Mestrom17}.
In theory, according to a general argument based on a variational principle applied to a finite range potential, the ground state trimer does not merge with the atom-dimer continuum ($a_\star^{(0)}$ does not exist)~\cite{Bruch73}.
It is also generally accepted that higher Efimov excited states ($n\geq2$) do merge and $a_\star^{(n)}$ are universally related to other Efimov features.
Hence, the behavior of the first excited Efimov state ($n=1$) is uncertain, especially in the vicinity of an intermediate Feshbach resonance ($s_{res}\approx1$).
Does it merge with the atom-dimer continuum in accordance with the $n\geq2$ excited states, or is there an avoided merger similar to the ground state despite the fact that the variational argument is not applicable?

This intriguing question, which is directly linked to the barely explored strength and origin of finite range corrections, cannot be addressed by currently available experimental techniques.
An apparent signature of merging of the first excited Efimov energy level with the atom-dimer continuum was obtained by measuring inelastic losses in an atom-dimer mixture~\cite{Knoop09,Bloom13,Zenesini14,Kato17}.
This method, however, is immune to the question of merging.
As was recently pointed out, enhancement in the atom-dimer inelastic collisional cross-section does not necessarily signify the merging point.
Instead, the trimer state could simply approach the atom-dimer continuum and then rebound from it, quickly becoming a deeply bound state~\cite{Mestrom17,Mestrom19}.
Another promising experimental method utilized loss spectroscopy to detect the trimer energy level directly, but failed to approach the merging point due to the unavoidable finite resolution limit set by the experimental conditions~\cite{Machtey12}.
Thus, known experimental techniques are no good for the task.

\section{Concept and Main Results}

Here we apply a recently developed experimental method to probe the first excited Efimov trimer in direct vicinity of its (possibly avoided) merger with the atom-dimer continuum (denoted $a_{\star}^{(1)}$, see Fig.~\ref{fig:illustration}a)~\cite{Yudkin19}.
The idea is to cast a system of three particles into a coherent dimer-trimer superposition (DITRIS) state and observe the accumulated phase difference of the quantum mechanical time evolution.
As a result we gain access to the relevant range of the Efimov energy level and demonstrate an order of magnitude improvement in precision compared to loss spectroscopy measurements~\cite{Lompe10RF,Nakajima11,Machtey12}.
In fact, previously applied methods for measuring three-body observables were predominantly based on inelastic (and mainly incoherent) scattering which causes loss of atoms.
In contrast, the experimental procedure applied here is lossless (the timescale is too short for particle loss) and potentially opens the door to study elastic scattering processes and the trimer's intrinsic lifetime.

Our main observation provides a definite answer: the first excited Efimov state does not merge with the atom-dimer continuum but instead rebounds from it.
With decreasing scattering length it approaches the continuum at first but then reemerges as a bound state to, most-likely, eventually become deeply bound.
Moreover, the functional form of the scattering length dependence of the binding energy around the point of minimal approach is highly non-universal which indicates the cross-over to a non-Efimovian three-body bound state.
In addition the magnetic moment of the bound state is observed and we show that it suddenly changes at the turning point.

\begin{figure*}
\centering
\includegraphics[width=0.8\linewidth]{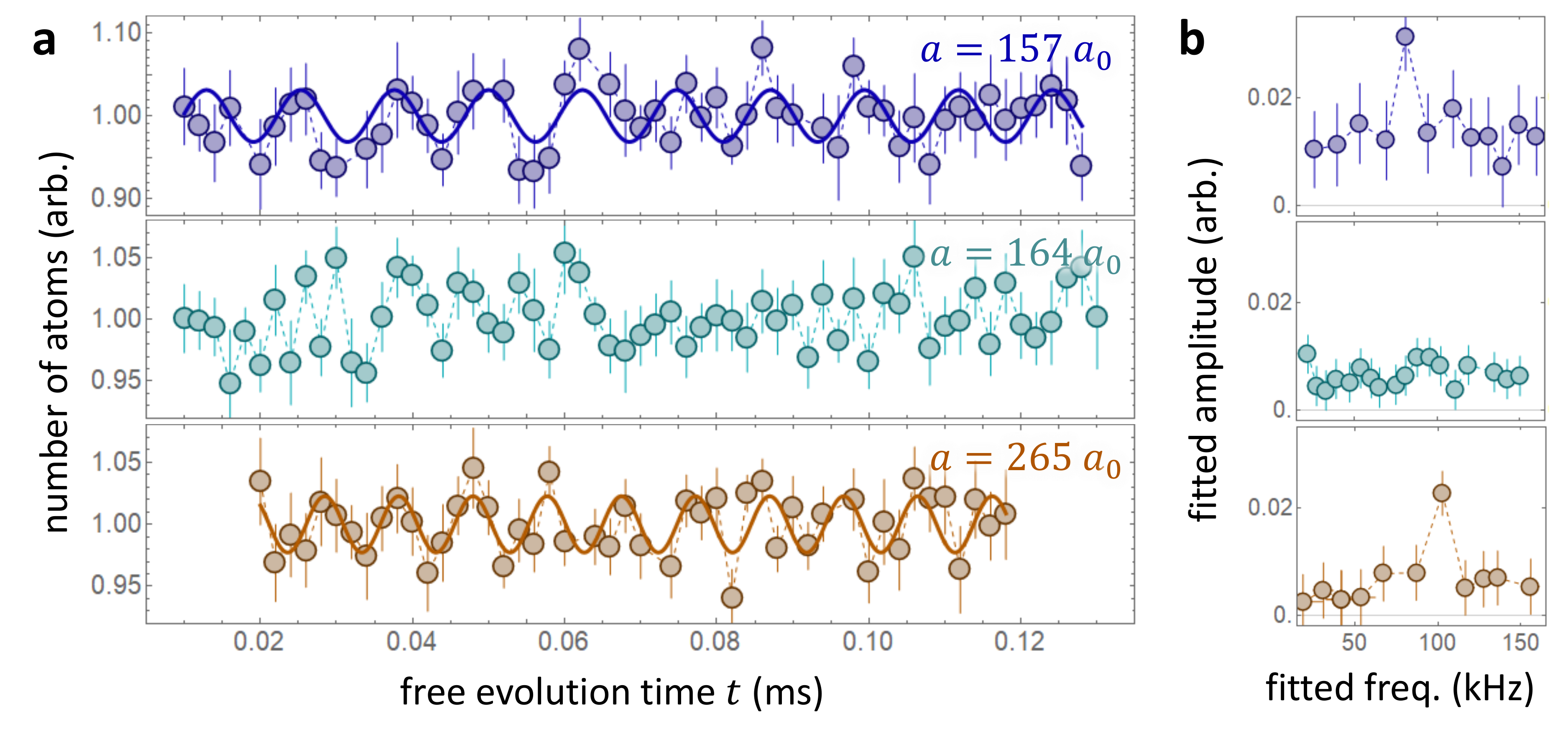}
\caption{\label{fig:number and spectrum illustration}
\textbf{Representative experimental data.}
{\bf a.}
Measurement of $N(t)$ for three different values of the scattering length.
Each point is the average of $10-20$ repetitions and the errorbar shows the standard deviation.
The solid curve in the upper and lower plot are the best fit to a sine.
{\bf b.}
Three-parameter fit of the $N(t)$ data in ({\bf a}).
The fitted frequency for which the fitted amplitude is maximal is used for the solid curves in ({\bf a}).}
\end{figure*}

\section{Experiment}

The DITRIS experiment is conducted on bosonic lithium ($^7$Li) atoms in the second lowest internal energy state which possesses two Feshbach resonances (see Appendix~\ref{ap:experimental details}).
The broader of the two is of intermediate strength between being open- and closed-channel dominated with $s_{res}=0.493$~\cite{Jachymski13}.
Based on the known position of $a_-^{(0)}=-280\;a_0$~\cite{Gross11} and applying universal relations one can na{\"i}vely predict $a_\star^{(1)}=288\;a_0$ ($a_0=0.529\times10^{-10}$ m is the Bohr radius).
However, experimental measurements at this scattering length showed that the trimer's binding energy remains $\gtrsim h\times100$ kHz below the atom-dimer continuum ($h=6.6\times10^{-34}$ m$^2$kg/s is Planck's constant)~\cite{Machtey12}.
In this experiment the binding energy was detected by radio-frequency (rf) loss spectroscopy based on inelastic (incoherent) collisions.
The resolution was limited by inherent power broadening of the dimer loss feature and denied access to trimer binding energies $|E_T|\lesssim|E_D|+h\times100$ kHz which prevented observation of the predicted trimer dissociation.
The coherent DITRIS experiment on the other hand is especially well-suited for the regime of interest: $|E_T-E_D|\lesssim100$ kHz~\cite{Yudkin19}.
As illustrated in Fig.~\ref{fig:illustration}c, a thermal gas of free atoms (temperature: $T=1.5\;\mu$K; $k_BT/h\approx30$ kHz, $k_B=1.38\times10^{-23}$ J/K is Boltzmann's constant) is subjected to a short ($10\;\mu$s) and strong rf modulation of the magnetic field.
The carrier frequency is $\sim E_D/h$ which falls in the range of $5-15$ MHz.
Note that $k_BT<|E_T-E_D|\ll E_D$, i.e. there is a clear separation of energy scales in our system (Fig.~\ref{fig:illustration}b).
Crucially, the modulation time is short enough so that its bandwidth covers both bound states (the dimer and the trimer) and therefore associates DITRIS states from the free atom continuum (Figs.~\ref{fig:illustration}a and~\ref{fig:illustration}c).
After a variable free evolution time $t$, during which each constituent of the superposition accumulates a phase proportional to its binding energy, a second pulse attempts to dissociate the molecules.
The time-dependent phase difference leads to interference in the final number of free atoms.
One expects the number of atoms $N(t)$ to vary sinusoidally with the free evolution time $t$ at the frequency $|E_T-E_D|/h$.
Measuring this oscillation is thus equivalent to measuring the trimer binding energy relative to that of the atom-dimer threshold.

In Fig.~\ref{fig:number and spectrum illustration}a we show the experimental $N(t)$ for a few representative values of the scattering length.
One notes the low signal-to-noise ratio (SNR) which is explained as follows.
The atoms are, to a good approximation, motionless during the pulse (because it is shorter than the inverse thermal energy $h/k_BT\approx30\;\mu$s) and hence DITRIS states are associated only if three atoms happen to be in close enough proximity to each other.
They don't have time to move around and look for partners.
This ultimately limits the DITRIS association efficiency, and hence the oscillation amplitude, to a few percent for the given experimental conditions.

The range of the free evolution time $t$ was chosen such that $N(t)$ is barely affected by the decoherence time $\tau$, caused either by elastic collisions or by the trimer's intrinsic lifetime.
The low SNR does not permit precise measurement of $\tau$ but empirically we do not observe signs of decay for $t<200\;\mu$s~\cite{Yudkin19}.
This deserves further in depth investigation, in practice however, by keeping $t<200\;\mu$s, we may neglect the decay in the data analysis~\cite{supMat}.

\section{Data Analysis and Discussion}

In order to detect the dominant frequency contribution we utilize a specially developed three-parameter fit analysis (3PA), inspired by Fourier transform but based on a least-squares fit to $N(t)/N_0=1+A\cos\left(\omega t+\varphi\right)$; $A$, $\omega$ and $\varphi$ being the fitting parameters and $N_0$ the average number of atoms~\cite{Yudkin19,supMat}.
Applied to $N(t)$ in Fig.~\ref{fig:number and spectrum illustration}a, the analysis results in the spectra shown in Fig.~\ref{fig:number and spectrum illustration}b.
The dominant frequency contribution for $265\;a_0$ ($157\;a_0$) is $102.6(9)$ kHz ($80.9(9)$ kHz).
The number in parentheses indicates 1$\sigma$ fitting errors.
We note that the value for $265\;a_0$ agrees within errorbars with our previous measurement for this scattering length, reported in~\cite{Yudkin19}.
For the intermediate value $164\;a_0$ no dominant oscillation frequency was detected.

Our method has two limitations: an upper and a lower bound.
The upper bound is due to the bandwidth of the pulse.
For too large $|E_T-E_D|/h$ the pulse no longer efficiently covers both bound states which diminishes the amplitude even more.
Considering the shape of the pulse we set our upper bound to $120$ kHz (see Appendix~\ref{ap:experimental details}).
The lower bound is connected to the inherent decay of coherence and the low SNR.
The 3PA needs at least four to five full periods to faithfully return the dominant frequency contribution.
Hence we are not able to detect frequencies below $20-25$ kHz.
Alternatively, if we assume a rather fast rate of elastic collisions, chemical equilibrium between trimers, dimers and free atoms results in a vanishing decay time for $|E_T-E_D|\lesssim k_BT$.
This gives a lower bound of $\sim30$ kHz.
Either way, the non-detected frequency at $164\;a_0$ is expected to be below this lower bound.

\begin{figure}[b]
\centering
\includegraphics[width=1.\linewidth]{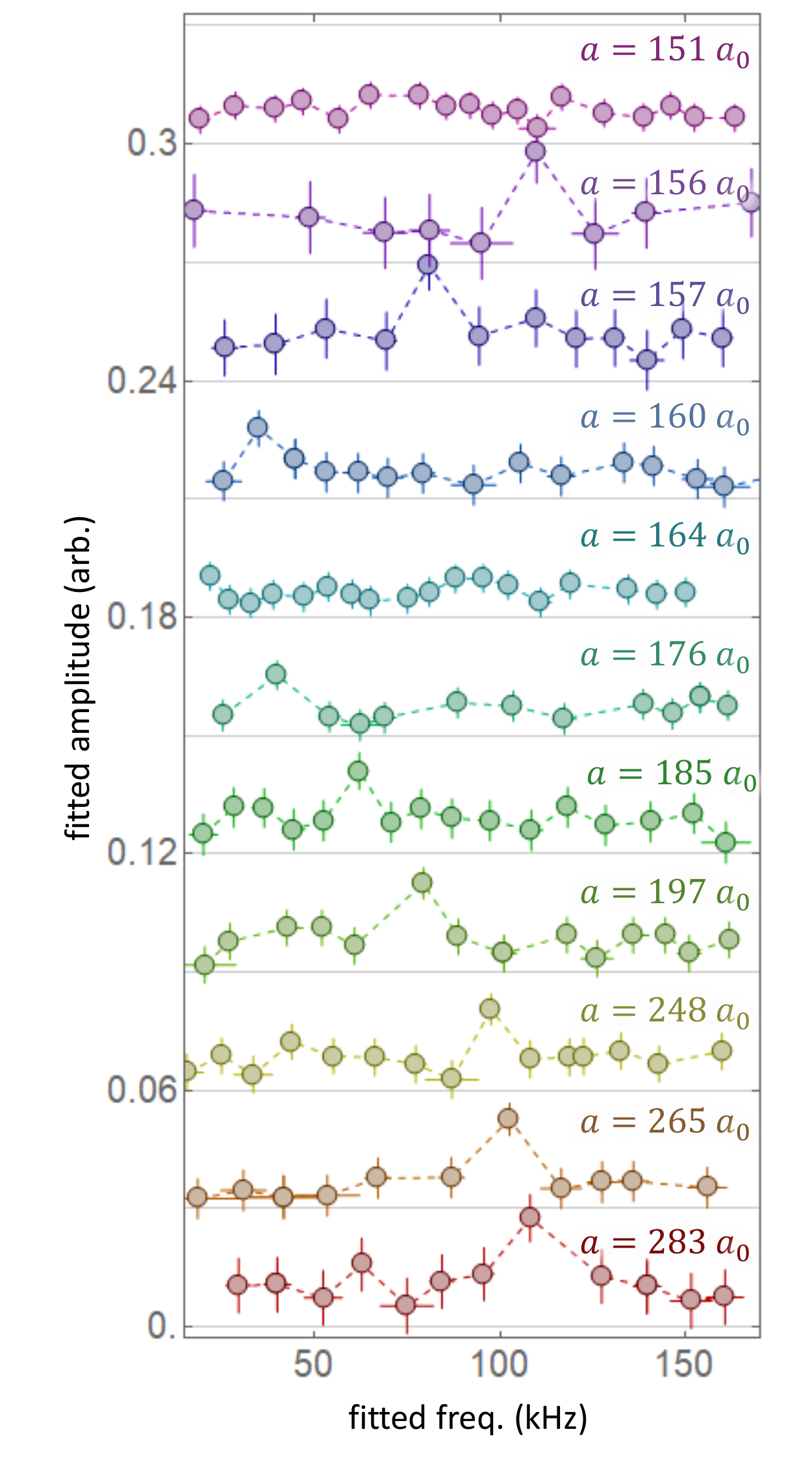}
\caption{\label{fig:spectrum}
\textbf{3PA for various $a$.}
The fitted amplitude is shown as a function of the fitted frequency for various values of the scattering length across the avoided merger.
The three spectra from Fig.~\ref{fig:number and spectrum illustration}b are shown again.
The curves are shifted with respect to each other by an offset for clarity.}
\end{figure}

\begin{figure*}
\centering
\includegraphics[width=0.8\linewidth]{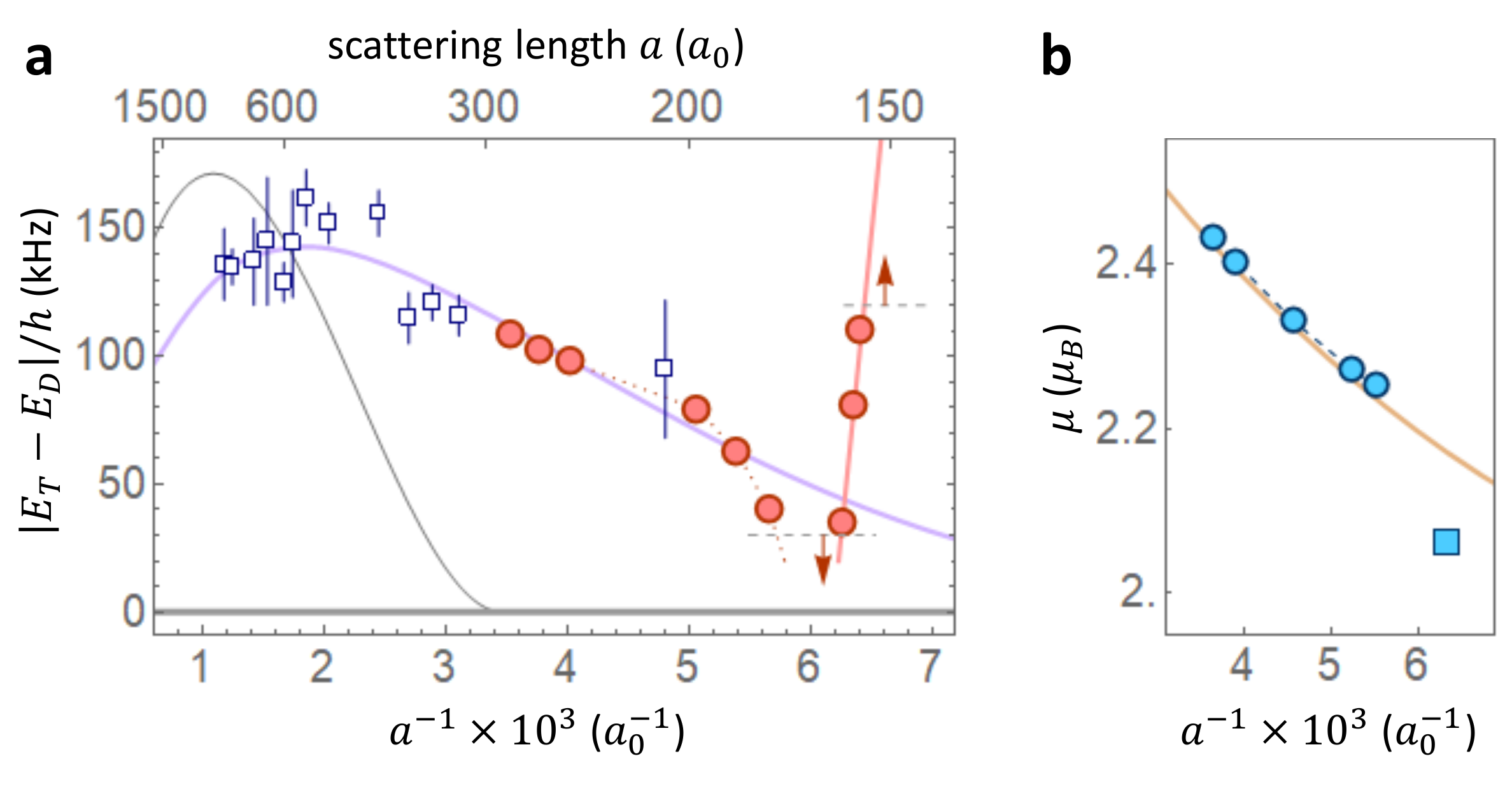}
\caption{\label{fig:ditris energy}
\textbf{Rebounding trimer.}
{\bf a.}
The dominant frequency contribution from Fig.~\ref{fig:spectrum} is plotted as a function of the inverse scattering length (red points).
The typical horizontal and vertical error bars, $\sim1a_0$ and $\sim1$ kHz respectively, are smaller than the point size.
The red arrows indicate the two values of $a$ at which no oscillations where detected and the dashed grey lines show our lower and upper detection limit.
The solid red line is a linear fit to the three points after the trimer rebounds from the atom-dimer continuum ($a<160a_0$).
For comparison we include the loss-spectroscopy measurements~\cite{Machtey12} as open blue squares, the universal theory as a grey line and the fit to an extended universal theory~\cite{Gattobigio14} as a purple curve.
{\bf b.}
Magnetic moment of the dimer + atom system (orange curve) and the trimer before (circles) and after (square) the avoided merger.}
\end{figure*}

To reveal how $|E_T-E_D|/h$ depends on $a$, and to witness the avoided merger, we have repeated the experiment at several values of $a$ (Fig.~\ref{fig:spectrum}).
For decreasing scattering length one sees that the dominant frequency decreases, vanishes, and then reappears and starts to increase rapidly.
This is also shown in Fig.~\ref{fig:ditris energy}a where the obtained data is plotted together with the rf loss spectroscopy data from~\cite{Machtey12}.
For comparison we also show the universal theory and a fit of the loss spectroscopy data to the universal functional form, extended to include finite range corrections~\cite{Gattobigio14} (see Appendix~\ref{ap:Efimov physics}).
In addition to $a/a_0=164$, also for $a/a_0=151$ no dominant frequency contribution was found by the 3PA (see Fig.~\ref{fig:spectrum} and red arrows in Fig.~\ref{fig:ditris energy}a); this time most probably due to the upper bound.

The grey line at $|E_T-E_D|=0$ in Fig.~\ref{fig:ditris energy}a shows the dissociation limit.
If it is reached the trimer dissociates into a dimer and a free atom (atom-dimer continuum).
Although our measurements seem to follow the finite-range corrected universal theory for $a/a_0>200$ (which predicts $a_*^{(1)}\approx80\, a_0$) they take a sudden turn towards the dissociation limit in stark disaccord with the theory.
It seems as if the dissociation limit will be reached when $a$ is further reduced to $\sim167\;a_0$.
This is supported by our failure to detect oscillations at $a/a_0=164$.
However, at $a/a_0<160$ the behavior changes abruptly and the trimer reemerges as a bound state.
Note that we cannot confirm that the trimer remains bound for all $160<a/a_0<176$.
It is possible that the trimer energy level turns around inside the continuum, where it would form a virtual bound state.

Either way, instead of a tangential and slow approach of the trimer energy level to the atom-dimer continuum (as predicted by the universal functional form) a rather narrow resonant feature is revealed in the experiment.
The behavior of $|E_T-E_D|$ thus raises the question: what finite-range effect (or effects) leads to the observed behavior?
Coincidental crossing of a non-universal three-body state is a possibility, however the fast change in the slope at $a/a_0\approx200$ is uncharacteristic for a standard avoided energy level crossing making this scenario implausible.

Although the question remains open it is instructive to follow the trimer's magnetic moment and identify a sudden change.
The slope of a molecules' binding energy, when plotted as a function of the magnetic field, is given by the magnetic moment of the molecule with respect to the free atoms.
Since our Feshbach resonance is located at high magnetic fields the electronic spins of the free atoms are nearly perfectly polarized (see Appendix~\ref{ap:experimental details}).
The magnetic moment of two free atoms is thus $2\mu_B$ ($\mu_B=1.4$ MHz/G is the Bohr magneton) while the deeply bound dimer is a pure singlet with zero magnetic moment.
For the magnetic field values relevant to our experiment the dimer is relatively shallow but deep enough to show non-universality (see Appendix~\ref{ap:experimental details}).
We extract its magnetic moment $\mu_D$, as a function of $a$, by performing a derivative of its binding energy $E_D$: $\mu_D=2\mu_B-\partial E_D/\partial B$.
Adding the third atom as a free atom (moment $\mu_B$) the polarized, three particle, dimer + atom system has $\mu_{DA}=\mu_D+\mu_B$.
This is plotted as a function of $1/a$ in Fig.~\ref{fig:ditris energy}b as a solid (orange) line.
The magnetic moment of the trimer is $\mu_T=3\mu_B-\partial E_T/\partial B$.
Applying a discrete derivative to our measurement of $E_T$ before it vanishes below the lower detection limit results in the circles in Fig.~\ref{fig:ditris energy}b.
The almost overlap with $\mu_{DA}$ indicates that the trimer is very similar in nature to the dimer + atom system.
However, the slope after rebounding (red solid line in Fig.~\ref{fig:ditris energy}a, $171$ kHz/G) has a stronger incline (and opposite sign) leading to a distinct and sudden change in $\mu_T$ (square in Fig.~\ref{fig:ditris energy}b).
We conclude that the nature of the trimer state changes, i.e. it becomes non-Efimovian, after it reemerges from below the detection limit.
This may also be inferred from the apparent functional form of the trimer energy level after it reemerges which is essentially a straight line (although this might be a local feature). 

A failure of the first excited Efimov state to merge with the atom-dimer continuum has been predicted in a recent single-channel theory, which is best-suited for open-channel dominated Feshbach resonances~\cite{Mestrom17}. 
It is found that the value of $a$ at which the trimer is closest to the dimer-atom continuum is $3.33\;r_{vdW}$ which for $^7$Li is $\approx103\;a_0$, substantially smaller than our measurement.
The avoided merger is interpreted to result mainly from a {\it d}-wave two-body resonance at $a\approx r_{vdW}$, inherent to the model.
In addition, at the point of closest approach a vanishing trimer life time is predicted contrary to our detection of a long-lived bound state (with respect to the oscillation period, i.e. $\tau\gg h/|E_T-E_D|$) in its vicinity.
A similar single-channel theory was able to tune the strength of the {\it d}-wave interactions and turn them off altogether~\cite{Mestrom19}.
It was shown that {\it d}-wave interactions enhance the avoided merger but it exists also in their absence.
As the Feshbach resonance used in the experiment is of intermediate character ($s_{res}\approx 1$) the above mentioned theories cannot be directly applied to our results.
In addition, $^7$Li does not have any {\it d}-wave resonances in the range $0-1200$ G (see Appendix~\ref{ap:experimental details}).
Therefore a different finite-range effect is responsible for our observation and its identification is a challenge for a future multi-channel theory.
A recent attempt at this was successful in describing Efimov features in cesium atoms ($s_{res}\gg1$) but failed to converge with the then available $^7$Li data~\cite{Wang14}.
Moreover, an avoided merger was not predicted.

\section{Conclusions and Outlook}

In conclusion, we have shown that, by exploiting coherence -- a quantum resource -- the Efimov trimer binding energy becomes experimentally accessible in a highly non-universal regime and can be measured with great accuracy.
This method was used to detect the non-crossing of the first excited Efimov state with the atom-dimer continuum and its reemergence as a non-Efimovian bound state.
Our observation challenges current theoretical models and may serve as precise and critical guide for new models.

An extension of our work is to measure the decay of the oscillations which are most-likely affected by temperature and density in addition to the more obvious scattering length dependence.
Such measurements rely on a considerable increase of the SNR which could be achieved by a different detection scheme sensitive to small atom numbers~\cite{Serwane11,Hume13}.
This would open the gate to study elastic scattering properties between atoms, dimers and trimers -- an observable which has so-far remained undetected.
Other atomic systems might also show avoided mergers and, hence, studying them is beneficial for deepening our understanding of finite-range effects.

\section*{Acknowledgements}
We are grateful to C. H. Greene for stimulating discussions and to P. S. Julienne for valuable clarifications on {\it d}-wave resonances of lithium and for providing coupled-channel calculations of the dimer binding energy and scattering length.
This research was supported in part by the Israel Science Foundation (Grant No. 1340/16).

\appendix

\section{Experimental Details}
\label{ap:experimental details}

\begin{figure*}
\centering
\includegraphics[width=0.7\linewidth]{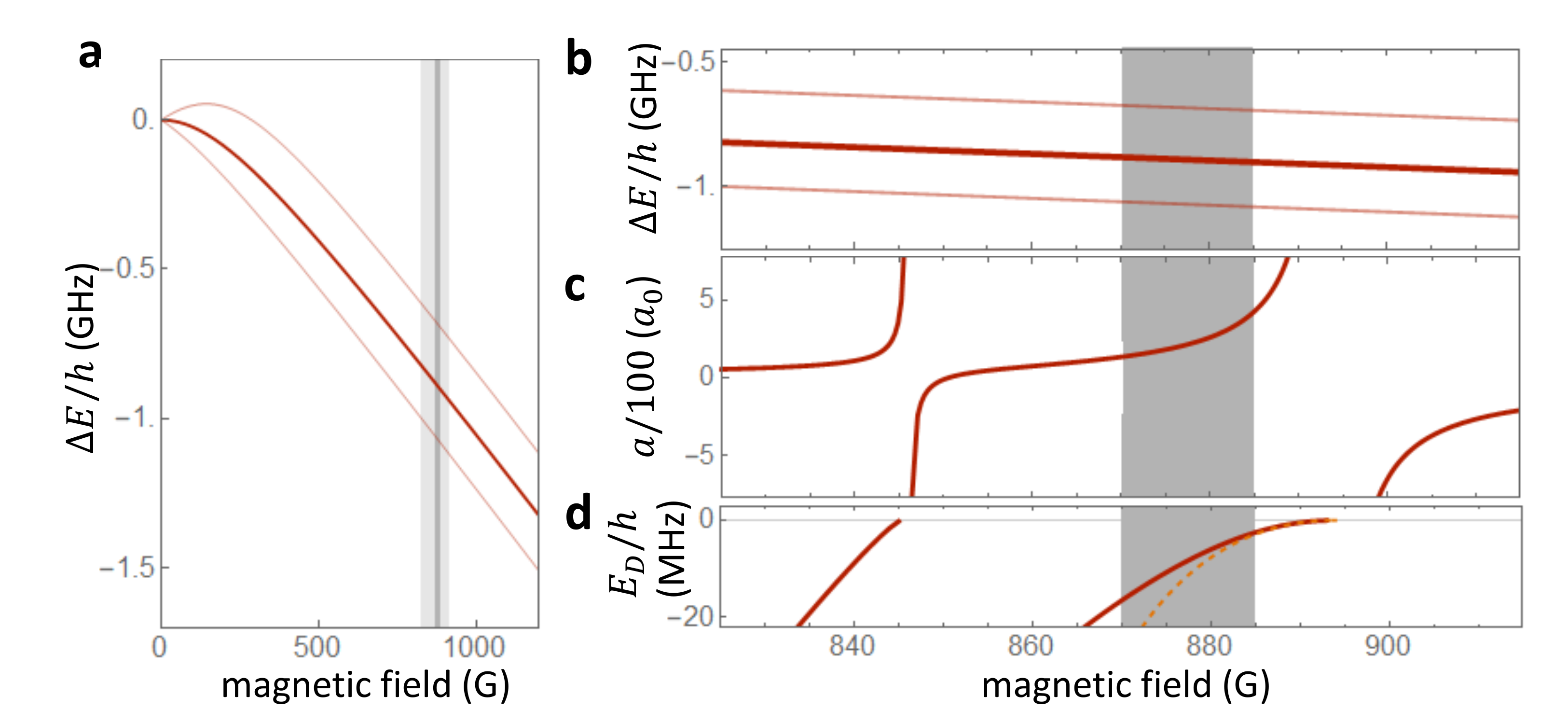}
\caption{\label{fig:f=1 manifold}
\textbf{The $F=1$ manifold of $^7$Li.}
{\bf a.}
Zeeman shift $\Delta E$ of the $m_F=1,0,-1$ (from bottom to top) sub-levels of $F=1$.
The light grey shading indicates the area enlarged in ({\bf b}).
The dark grey shaded region, here and in ({\bf b}-{\bf d}), marks the magnetic field values explored in this work.
{\bf b.}
Enlarged view of ({\bf a}).
{\bf c.}
Scattering length vs magnetic field.
A narrow and an intermediate Feshbach resonance are visible.
{\bf d.}
Feshbach dimer binding energies for the two resonances.
The orange dashed curve is the universal prediction $E_D\sim-1/a^2$.}
\end{figure*}

Standard laser cooling and evaporation techniques are used to produce a gas of $3\times10^4$ bosonic lithium atoms at $1.5\;\mu$K and an average density of $1.25\times10^{12}$ cm$^{-3}$ in a crossed optical dipole trap.
The work in performed in the lower hyper-fine manifold ($F=1$) of the ground state whose Zeeman sub-levels are shown in Fig.~\ref{fig:f=1 manifold}a as a function of the magnetic field.
Within this manifold all atoms are pumped at the $m_F=0$ Zeeman sub-level.
One notes that for the relevant field values the Zeeman shift is linear (see also Fig. ~\ref{fig:f=1 manifold}b) demonstrating that the gas is polarized (all electronic spins point along the direction of the field) and hence two free atoms interact through a triplet potential.
The magnetic field dependence of the scattering length $a$ is shown in Fig.~\ref{fig:f=1 manifold}c for the $m_F=0$ state.
The Feshbach resonance at $846$ G is closed-channel dominated ($s_{res}\ll1$) while the resonance at $893.7$ G, in whose vicinity our work is conducted, is intermediate ($s_{res}=0.493$).
Both are {\it s}-wave resonances.
Lithium does not feature any higher partial wave resonances for fields below $1200$ G.
The binding energy of the Feshbach molecules is shown for both resonances in Fig.~\ref{fig:f=1 manifold}d.
In the regime of interest the binding energy is different from the universal dimer $-\hbar^2/ma^2$, where $m$ is the mass of a single lithium atom.
In particular, at deep binding energies (not shown) the dimer becomes a pure singlet and thus has a linear magnetic field dependence with respect to the free atom trimer state.

\begin{figure}
\centering
\includegraphics[width=1.\linewidth]{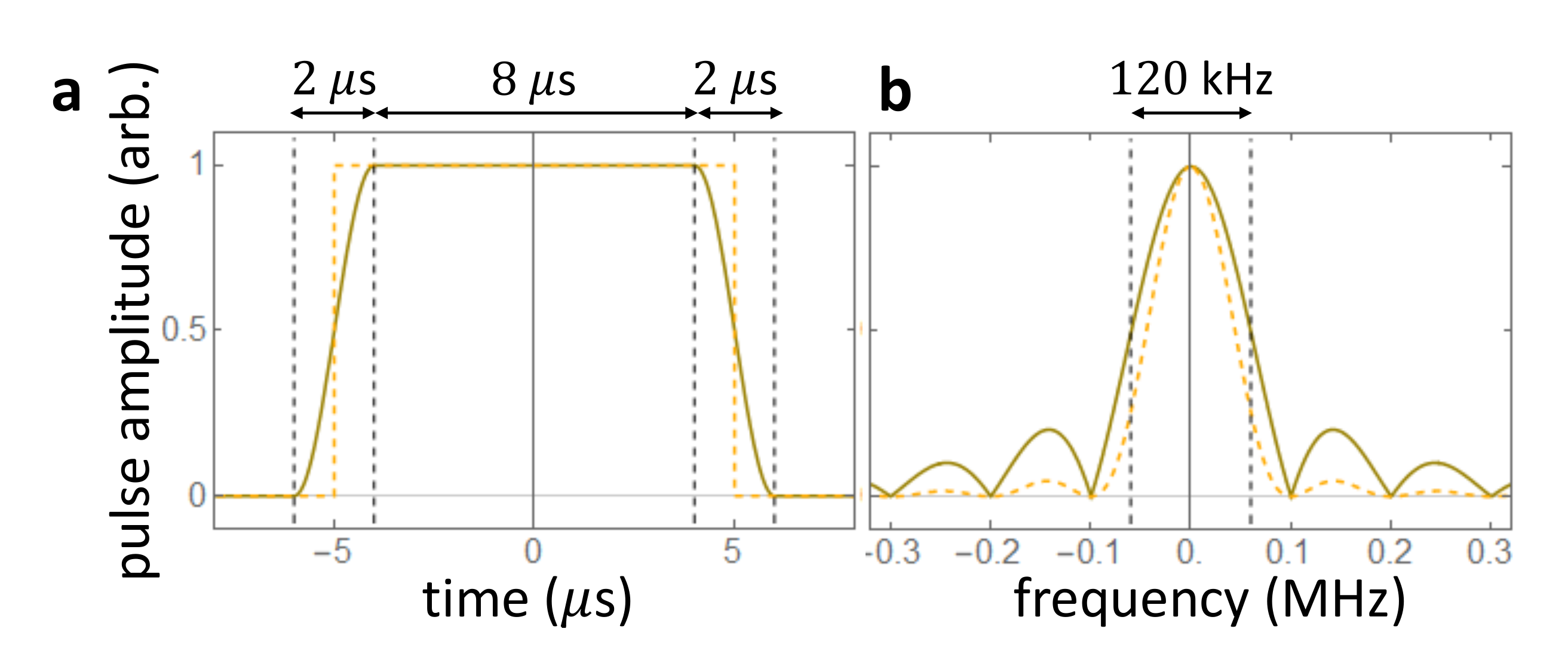}
\caption{\label{fig:pulse}
\textbf{Pulse.}
{\bf a.}
Plot of the pulse shape in Eq.~(\ref{eq:pulse shape}) (solid) and ideal rectangular pulse (dashed).
{\bf b.}
Fourier transform of ({\bf a}).}
\end{figure}

At the core of the experiment lies the $10\;\mu$s pulse which is Fourier broadened to address both the dimer and the trimer simultaneously.
The duration $10\;\mu$s refers to the full-width at half-maximum (FWHM).
There is also a (measured) turn-on/turn-off time of $\tau_0=2\;\mu$s which means that the pulse is at its maximal value during $\tau_c=8\;\mu$s.
The experimental rf pulse envelope is modeled as (Fig.~\ref{fig:pulse}a):
\begin{equation}
\chi(t)=\begin{cases}
\sin^2\left(\frac{\pi}{2}\frac{t+\tau_c/2+\tau_0}{\tau_0}\right) & -\tau_0-\tau_c/2<t<-\tau_c/2 \\
1 & -\tau_c/2<t<\tau_c/2 \\
\sin^2\left(\frac{\pi}{2}\frac{t-\tau_c/2-\tau_0}{\tau_0}\right) & \tau_c/2<t<\tau_c/2+\tau_0
\end{cases}
\label{eq:pulse shape}
\end{equation}
The Fourier transform of $\chi(t)$ is shown in Fig.~\ref{fig:pulse}b in the low frequency domain.
It closely resembles a sinc, the transform of an ideal rectangular pulse, but is slightly broadened.
It's FWHM is $120$ kHz which we use as our upper detection limit.

\section{Short Summary of the Relevant Efimov Physics in Bosonic Lithium}
\label{ap:Efimov physics}

In the $m_F=0$ channel of $^7$Li, the intersection of the ground state Efimov energy level with the three-atom continuum was found to be at $a_{-}^{(0)}=-280\;a_0$~\cite{Gross11}.
Universal theory then predicts that the $a\rightarrow\infty$ binding energy of the first excited Efimov trimer is
\begin{equation}
E_T^{(1)}\left(\frac{1}{a}=0\right)=-\frac{\hbar^2(\kappa_{\star}^{(1)})^2}{m}=-h\times31.1\text{ kHz},
\end{equation}
where $\kappa_{\star}^{(1)}=-1.56/a_{-}^{(1)}$ and $a_{-}^{(1)}=a_{-}^{(0)}e^{\pi/s_0}$ ($s_0=1.00624$).
In addition, the trimer is expected to merge with the atom-dimer continuum (represented by the universal dimer energy level $E_D=-\hbar^2/ma^2$) at $a_{\star}^{(1)}=0.07/\kappa_{\star}^{(1)}=288\;a_0$ which corresponds to a dimer binding energy of $-h\times6.21$ MHz.

In the framework of zero range universal theory, given $\kappa_{\star}^{(n)}=e^{-\pi n/s_0}\kappa_{\star}^{(0)}$, the binding energy of the $n$-th excited state is found by solving
\begin{equation}
\frac{E_T^{(n)}}{E_D}=\tan^2\xi ,\quad \kappa_\star^{(n)}a=\frac{1}{h(\xi)\cos\xi},
\end{equation}
where $h(\xi)=e^{\Delta(\xi)/2s_0}$ and $\Delta(\xi)$ was found numerically and approximated by analytic expressions in~\cite{Ananos03} and improved by~\cite{Naidon17}.
The angle $\xi$ takes values from $-\pi$ to $-\pi/4$.
Note that the vertical axis $1/a=0$ is given by $\xi=-\pi/2$ and the region $a>0$ discussed in this paper by $-\pi/2<\xi<-\pi/4$.
In Fig.~\ref{fig:ditris energy}a we have plotted $|E_T^{(1)}-E_D|$ as a function of $1/a$ for $a>0$ as a grey line.

Evidently, the loss-spectroscopy data disputes the universal theory (blue squares in Fig.~\ref{fig:ditris energy}a)~\cite{Machtey12}.
A possible way to extend the universal theory to include finite range corrections has been suggested in~\cite{Gattobigio14}.
The idea is to use the exact dimer binding energy instead of the universal relation, thus defining an effective scattering length $a_B$ such that $E_D=-\hbar^2/ma_B^2$ (even though or since $E_D\neq-\hbar^2/ma^2$), and to rescale and shift the three-body parameter: $\kappa_{\star}a\rightarrow\kappa_3a_B+\Gamma_3$.
The binding energy of the $n$-th trimer is given by
\begin{equation}
\frac{E_T^{(n)}}{E_D}=\tan^2\xi ,\quad \kappa_3^{(n)}a_B+\Gamma_3^{(n)}=\frac{1}{h(\xi)\cos\xi}.
\end{equation}
By fitting the model to the experimental data ($n=1$) one finds $\kappa_3^{(1)}=1.61\times10^{-4}/a_0$ and $\Gamma_3^{(n)}=4.95\times10^{-2}$~\cite{Gattobigio14}.
This result is shown as a purple line in Fig.~\ref{fig:ditris energy}a.


\clearpage
\onecolumngrid


\setcounter{figure}{0}
\setcounter{equation}{0}
\renewcommand\thefigure{S\arabic{figure}}
\renewcommand\theequation{S\arabic{equation}}

\begin{center}
{\Huge Supplemental Material}
\vspace{0.5cm}
\end{center}

Here we elaborate on the idea and procedure of the three-parameter fit analysis (3PA) used to extract the dominant frequency contribution from our low-SNR data.
Please also refer to the Supplemental Material of~\cite{Yudkin19}, where this method was first introduced.

\section*{Detailed Description of the Three-Parameter Analysis}

\begin{figure*}[b]
\centering
\includegraphics[width=1.\linewidth]{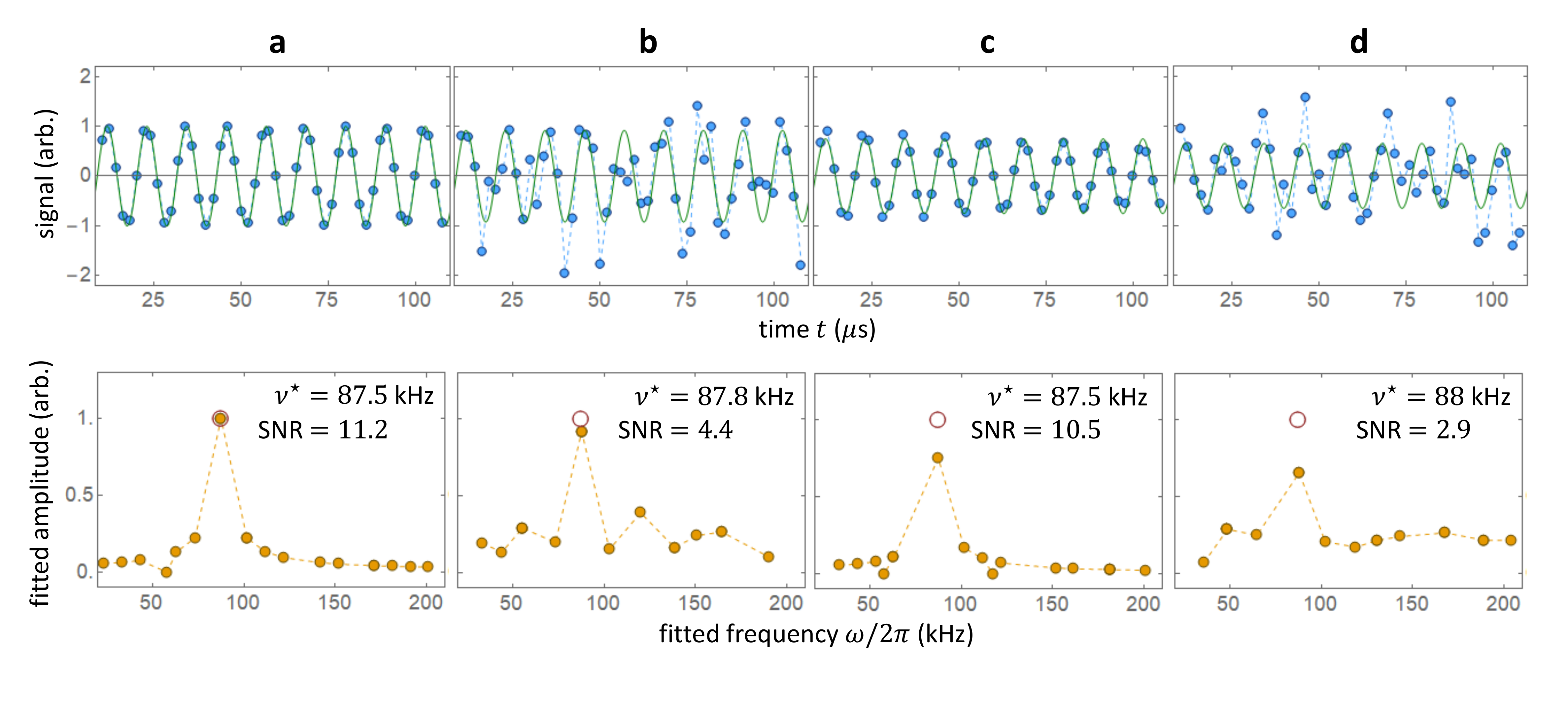}
\caption{\label{fig:details of 3PAR}
\textbf{3PA of a simulated signal.}
{\bf a.}
Time- (upper row) and frequency-domain (lower row) of a sinusoidal signal without noise.
{\bf b.}
Same as ({\bf a}), with noise.
{\bf c.}
Same as ({\bf a}) for decaying oscillations, no noise.
{\bf d.}
Same as ({\bf c}), with noise.
}
\end{figure*}

In order to illustrate our data analysis it is instructive to apply it to a simulated data sequence.
Consider a finite-length sinusoidal signal similar to the one shown in the upper two of Fig.~\ref{fig:details of 3PAR}a, for which a $100\;\mu$s long pure sine with $\omega/2\pi=87.5$ kHz was generated.
As in our experiment a discrete ``measurement'' value is taken every $2\;\mu$s corresponding to a sampling rate of $500$ kHz.
In order to determine the frequency we guess a pure oscillatory fitting function:
\begin{equation}
N(t)=A\cos(\omega t+\varphi).
\label{eq:cosine fitting function}
\end{equation}
The three fitting parameters are the amplitude $A$, the frequency $\omega$ and the phase $\varphi$.
Since the frequency is not known a priori we instruct the least-squares algorithm to start its search for a minimum in parameter space $(A,\omega,\varphi)=(1,\omega_0,0)$, where $\omega_0 \in 2\pi\times[20, 200]$ kHz.
For each initial value of $\omega_0$ the algorithm converges to some value for the three parameters $(A,\omega,\varphi)$ in the vicinity of the initial parameters (possibly a local minimum, not the global minimum) and we record the converged $A(\omega)$ (Fig.~\ref{fig:details of 3PAR}a, lower row).
The value of $\omega$ at which $A$ is maximal (we denote these values $A^\star$ and $\omega^\star$) is the dominant frequency contribution and the global minimum in parameter space.
As expected for this pure sine, $\omega^\star/2\pi=87.5$ kHz is obtained.
The trustworthiness of the spectrum is quantified by a signal-to-noise ratio as:
\begin{equation}
\text{SNR}=\frac{A^\star}{\bar{A}},
\end{equation}
where $\bar{A}$ is the mean of all points excluding $A^\star$.
For the pure sine, SNR $=11.2$.
Due to the finite length of the signal, $\bar{A}\neq0$ and hence the SNR does not diverge.

We now add white Gaussian noise (WGN) with a standard deviation of $0.5$ (half the amplitude) to the pure sine and repeat the procedure (Fig.~\ref{fig:details of 3PAR}b).
Albeit the WGN, the 3PA is able to determine the dominant frequency contribution with an error $<1kHz$ (corresponding to the typical errorbar) but with a reduced SNR $=4.4$.
Note that it is not $A^\star$ that is lowered due to the WGN but $\bar{A}$ which is increased.

The real signal of the DITRIS experiment decays.
A decaying sine with characteristic decay time $\tau=200\;\mu$s is simulated without noise in Fig.~\ref{fig:details of 3PAR}c.
One notes that the obtained $\nu^\star$ is identical to the non-decaying signal of Fig.~\ref{fig:details of 3PAR}a although Eq.~(\ref{eq:cosine fitting function}) was used to determine it and that the SNR is reduced by less than $10\%$

Finally, Fig.~\ref{fig:details of 3PAR}d shows a noisy decaying sine representing the real experimental conditions.
Despite the fitting function not including the decay and despite the noise, the frequency $\nu^\star$ is found with an error $<1$ kHz!
Although the SNR is reduced by a factor of $\sim3$ with respect to Fig.~\ref{fig:details of 3PAR}a the dominant frequency contribution is easy to read off the lower row of Fig.~\ref{fig:details of 3PAR}d.

\section*{Alternative Analyses}

The 3PA is better than a fast Fourier transform (FFT).
The main reason for this is the finite sample length.
For the signals in Fig.~\ref{fig:details of 3PAR} the sample length ($100\;\mu$s) leads to a frequency resolution of $(100\;\mu$s$)^{-1}=10$ kHz (irrelevant of the sampling rate).
The accuracy of frequency determination is thus limited to $10$ kHz.
In our case, where $87.5$ kHz is the correct frequency, both the $80$ kHz and the $90$ kHz point have an amplitude of $\sim0.5$, heavily reducing the accuracy and the SNR.
The FFT method is very well suited for long samples but not for our relatively short data.

\begin{figure*}
\centering
\includegraphics[width=0.8\linewidth]{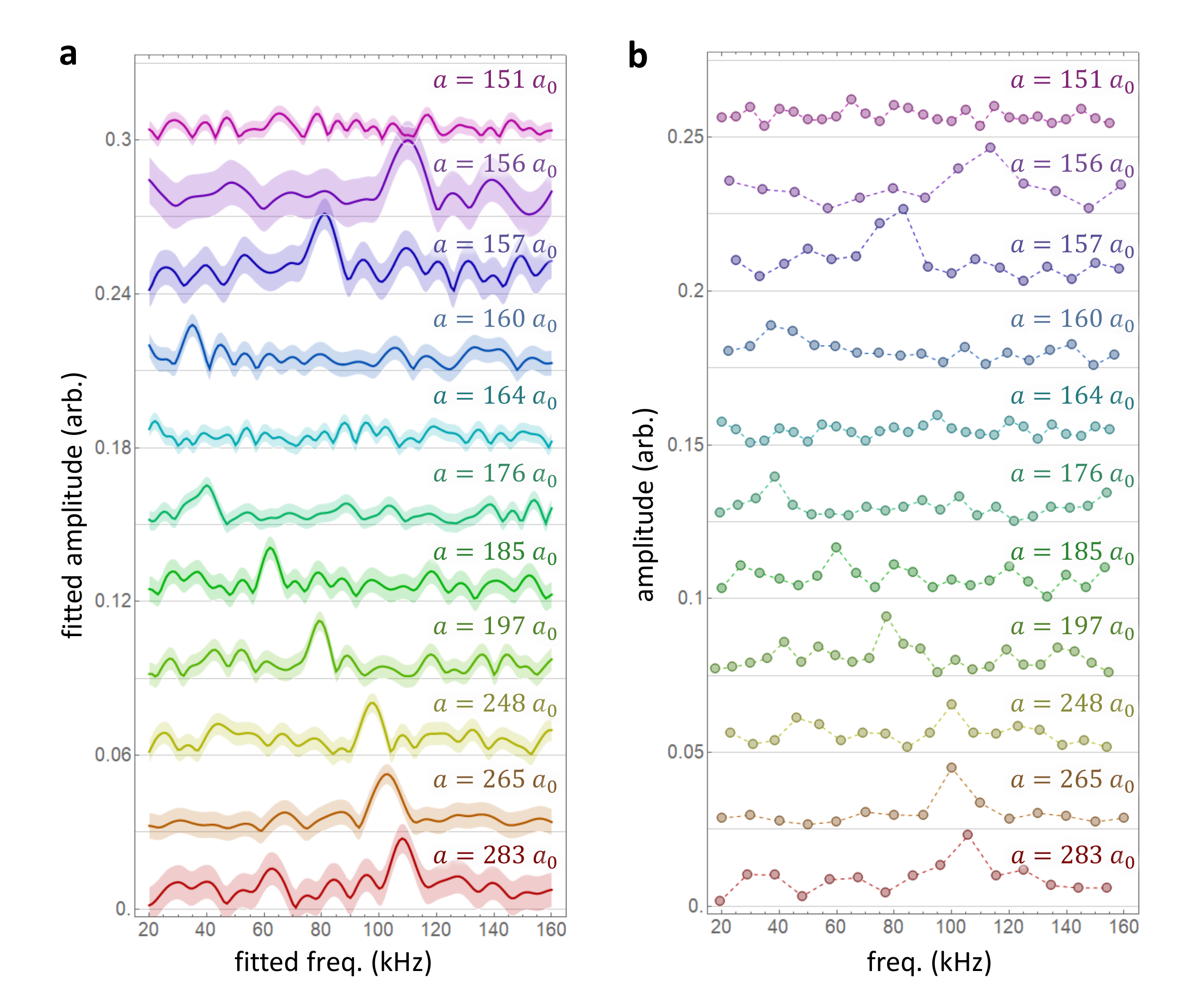}
\caption{\label{fig:spectrum supplement}
\textbf{2PA and FFT.}
({\bf a.})
2PA of data.
The solid line marks the best fit and the shaded region is 1$\sigma$ fitting error.
({\bf b.})
FFT of data.}
\end{figure*}

An alternative to the 3PA is a two-parameter fit analysis (2PA).
This involves a fit to Eq.~(\ref{eq:cosine fitting function}) but using only $A$ and $\varphi$ as fitting parameters.
The frequency is put in by hand and the least-squares algorithm finds the best fitting amplitude and phase for each frequency.
Although this method does not suffer from finite resolution, which may be made arbitrarily small, the likelihood analysis, described next, shows its clear disadvantage.
In addition, fixing $\omega$ does not provide fitting errors for the frequency.
The 3PA on the other hand provides a confidence interval for all three parameters.
Figs.~\ref{fig:number and spectrum illustration}b and~\ref{fig:spectrum} of the main text show these as errorbars.

For completeness and in addition to the 3PA we have analyzed our experimental $N(t)$ with the 2PA and FFT (Fig.~\ref{fig:spectrum supplement}).
Unsurprisingly the same signature as in Fig.~\ref{fig:spectrum} of the main text is obtained.

A comparison of all three methods (FFT, 2PA and 3PA) can be found in the Supplemental Material of~\cite{Yudkin19}.

\section*{Likelihood Analysis}

Here we show that 3PA is the least likely to be fooled by a false signal.
The question we answer here is:
For a sample of random numbers, how likely is the 3PA to find a dominant frequency contribution even though none is there (a so-called false-positive)?

\begin{figure}[b]
\centering
\includegraphics[width=0.4\linewidth]{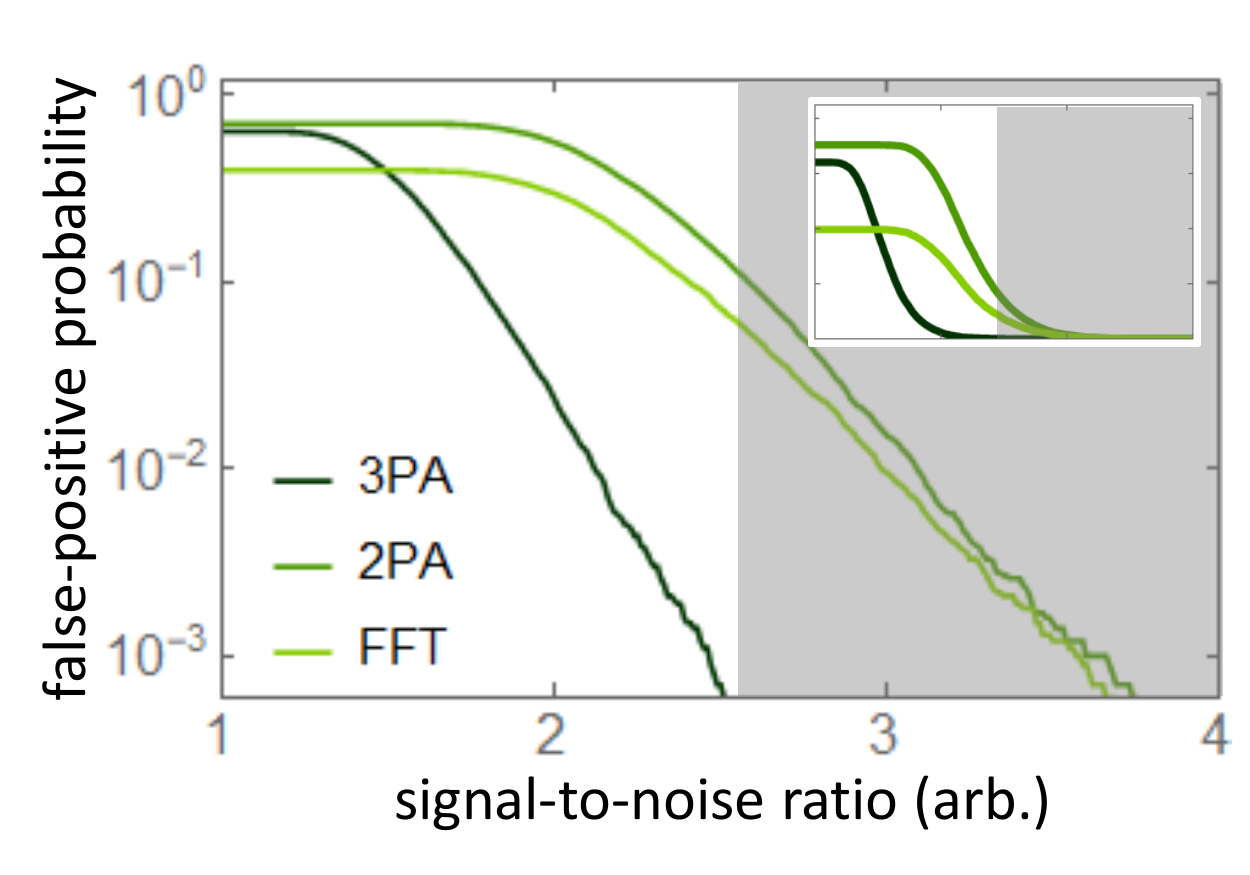}
\caption{\label{fig:likelihood}
\textbf{Likelihood analysis.}
The probability of detecting a false-positive is plotted vs the signal-to-noise ratio for three different analysis methods.
The grey shaded region indicates the experimental SNR values.
The inset shows the same data on a linear scale.
}
\end{figure}

To this end we generate $10^4$ fake signals, each $100\;\mu$s long and sampled at a $500$ kHz rate.
The random numbers are drawn from a Gaussian random number generator with $0.035$ standard deviation (derived from the experimental signals similar to those in Fig.~\ref{fig:number and spectrum illustration}a of the main text).
We run all three analysis methods on each signal and, as a function of SNR $=A^\star/\bar{A}$, count the number of false-positives.
The result, presented in Fig.~\ref{fig:likelihood}, shows that for $SNR>1.6$ the 3PA is least likely to be fooled by a false positive and that for SNR $>2.47$ this probability drops below the $10^{-3}$ level .
The other two methods require an SNR of $3.67$ (2PA) and $3.59$ (FFT) respectively to obtain the same probability.

None of the curves reaches unity for SNR $\leq 1$ because we only consider frequency values within our physically accessible window $30$ kHz $<\omega^\star/2\pi<150$ kHz (bounded by finite resolution limits discussed in the main text).
Especially the FFT finds mainly high frequencies.
The range of our experimental SNR values is indicated by the grey region in Fig.~\ref{fig:likelihood}.
Since the false-positive probability of the 3PA is the lowest in this region we consider it as the most reliable method to analyze the data.

\section*{Statistical Significance of Presented Results}

To clarify the statistical significance of our results we recall the measurement presented in~\cite{Yudkin19}.
There, the $a/a_{0} = 265$ point was measured with SNR = $2.79(55)$ and the reported frequency was $101.9(8)$~kHz.
In this work we returned to the same value of the scattering length and extracted the frequency $102.6(9)$~kHz with improved SNR = $3.74(45)$.
The two frequencies agree with each other within 1$\sigma$. 
Based on our likelihood analysis, the probability to obtain this coincidence by chance drops significantly below $10^{-6}$ (equivalent to a statistical significance of more than $5\sigma$).
Finally, the observation of a meaningful scattering length dependence of the extracted frequencies at $11$ distinguished values makes this discussion essentially redundant.

\section*{Cramer-Rao Lower Bound}

When considering low-SNR sinusoidal data the following question naturally arises:
How much could we benefit from increasing the sampling rate at the expense of shortening the sample length?
In other words, given a finite number of data points, is it better to spread them out over many oscillations or to sample the first oscillation very densely?

To answer this question we look at the standard signal-processing procedure called Cramer-Rao lower bound (CRLB)~\cite{Kay}.
Consider $N$ samples obtained at times $t_n$ ($n=1,\dots,N$):
\begin{equation}
x_n=A\cos\left(\omega t_n+\varphi\right)+w_n(0,\sigma),
\end{equation}
where $w_n$ is a Gaussian random number with zero mean and standard deviation $\sigma$.
The values of $A$, $\omega$ and $\varphi$ are not known.
The CRLB provides a mathematical expression for how well their value may be estimated for a given $\sigma$ and $N$.
In the following we consider two cases:
\begin{itemize}
\item
$t_n=(n-1)dt$
\item
$t_n=(n-1)dt/N$
\end{itemize}
In the first case, increasing $N$ makes the sample longer but the sampling rate ($1/dt$) remains constant.
For the second option the opposite is the case.
To find the CRLB we need the probability distribution function (PDF) of the $n$-th point:
\begin{equation}
p_n\left(x_n;A,\omega,\varphi\right)=\frac{1}{\sqrt{2\pi\sigma^2}}\exp\left[-\frac{\left(x_n-A\sin(\omega t_n+\varphi)\right)^2}{2\sigma^2}\right].
\end{equation}
The PDF of the entire data set $x=\{x_n\}$ is $p\left(x;A,\omega,\varphi\right)=\prod_np_n$.
The CRLB theorem claims that the lower bound for estimating $A$, $\omega$ or $\varphi$ is given by the inverse of the curvature of $p\left(x;A,\omega,\varphi\right)$ in parameter space (spanned by $A$, $\omega$ and $\varphi$).
The curvature, moreover, is given by the negative of the second log derivative.
The lower bound is thus computed in two steps.
First we must arrange all second partial derivatives of $\ln[p\left(x;A,\omega,\varphi\right)]$ into a matrix known as the Fisher information matrix:
\begin{equation}
\hat{F}=\begin{pmatrix}
-\frac{\partial^2\ln p}{\partial A^2} & -\frac{\partial^2\ln p}{\partial A \partial \omega} & -\frac{\partial^2\ln p}{\partial A \partial \varphi} \\
-\frac{\partial^2\ln p}{\partial \omega \partial A} & -\frac{\partial^2\ln p}{\partial \omega^2} & -\frac{\partial^2\ln p}{\partial \omega \partial \varphi} \\
-\frac{\partial^2\ln p}{\partial \varphi \partial A} & -\frac{\partial^2\ln p}{\partial \varphi \partial \omega} & -\frac{\partial^2\ln p}{\partial \varphi^2}
\end{pmatrix}.
\end{equation}
Note that for any element of $\hat{F}$ that depends explicitly on $x_n$ the expectation value weighted by $p\left(x;A,\omega,\varphi\right)$ must be taken.
In the second step we compute the inverse matrix and keep the on-diagonal elements.
The lower bound variance of the $i$-th parameter estimation is given by $(\hat{F}^{-1})_{ii}$.

We have numerically computed this value as a function of $N$ in both cases outlined above.
The frequency lower bound var$(\omega)\geq(\hat{F}^{-1})_{22}$ is found to be $\sim1/N^3$ in the first case (increasing $N$ means increasing sample length) and only $\sim1/N$ in the second (increasing $N$ means increasing sampling rate).
By increasing the sample length one thus benefits from an additional factor of $1/N$ (note that the variance is the square of the standard error).
For frequency estimation it is thus advantageous to sample at a low rate and for a long time.

In our experiment the sample length is ultimately limited by the decay of the signal which, as discussed in the main text, is $>200\mu$s.

\end{document}